\documentclass[letterpaper, 10 pt, conference]{ieeeconf}  

\IEEEoverridecommandlockouts                              

\usepackage[utf8x]{inputenc}
\usepackage{etoolbox}
\usepackage{amsmath}
\usepackage{amssymb}
\usepackage{comment}
\usepackage{hyperref}
\usepackage{breakurl}
\usepackage{caption}
\usepackage{subcaption}
\usepackage{color}
\usepackage{textcomp}
\usepackage{graphicx}      
\usepackage{cite}
\usepackage{url}
\usepackage[font=small,labelfont=sf,justification=justified,format=plain]{caption}
\usepackage{array}
\usepackage{makecell}
\usepackage[export]{adjustbox}

\newtheorem{myprob}{Problem}

{
\hyphenation{op-tical net-works semi-conduc-tor}

\title{\LARGE \bf
Design of a Smooth Landing Trajectory Tracking System for a Fixed-wing Aircraft
}
\author{Solomon Gudeta and  Ali Karimoddini
\thanks{S. G. Gudeta and A. Karimmodini are with the Department of Electrical and Computer Engineering, North Carolina Agricultural and Technical State University, Greensboro, North Carolina, USA.}
\thanks{Corresponding author: A. Karimoddini, Tel: +13362853313, {\tt\small akarimod@ncat.edu}.}
}

\begin{document}
\maketitle
\thispagestyle{empty}
\pagestyle{empty}

\begin{abstract}
This paper presents a  landing controller for a fixed-wing aircraft during the landing phase, ensuring the aircraft reaches the touchdown point smoothly. The landing problem is converted to a finite-time linear quadratic tracking (LQT) problem in which an aircraft needs to track desired landing path in the longitudinal-vertical plane while satisfying performance requirements and flight constraints. First, we design a smooth trajectory that meets flight performance requirements and constraints. Then, an optimal controller is designed to minimize the tracking error, while landing the aircraft within the desired time-frame. For this purpose, a linearized model of an aircraft developed used under the assumption of a constant approach airspeed, is used. The resulting Differential Riccati equation is solved backward in time using the Dormand Prince algorithm. Simulation results show a good tracking performance and the finite-time convergence of tracking errors for different initial conditions of the flare-out phase of landing.


\end{abstract}


%

\section{Introduction}
One of the most difficult tasks in aircraft control is achieving a safe and comfortable landing. A review of accident statistics indicates that over 45 percent of all general aviation accidents occur during the approach and landing phases of a flight \cite{aviation2004airplane}. A closer look shows that the cause of over 90 percent of those cases was pilot related. Loss of control was also a major contributing factor in 33 percent of these cases. In many cases, the accidents have happened due to not landing within a targeted landing time \cite{ellert1963synthesis}. Aircraft landing has fives phases: the base leg, the glide slope, the flare-out, the touchdown, and the after-landing roll. The glide slope and the flare-out are the most crucial phases. In the glide slope phase the aircraft descends along a predefined straight line at a steady state descent angle; whereas, in the flare-out phase the aircraft gradually raises its nose while landing. There are several performance requirements and constraints associated with aircraft landing. The aircraft needs to follow the desired trajectory and land smoothly within a targeted landing time.  In addition, the aircraft landing controller should be robust to wind turbulence, measurement noise, and actuator failures.

Many researchers have applied various control techniques to design an automatic aircraft landing control system. Different control methods such as classical control \cite{juang2008hardware, tripathi2018autonomous}, optimal control \cite{ellert1963synthesis,shue1999design,lungu2015application,wu2015take}, adaptive control \cite{rao2014automatic,wagner2007digital}, nonlinear control \cite{tripathi2016autonomous,prasad2007automatic}, and intelligent control \cite{lungu2017landing,lungu2013automatic, lau2007neural} have been used to solve the automatic aircraft landing problem. In \cite{stevens2015aircraft}, the flare-out maneuver control was successfully implemented using the classical feedback control method. \cite{juang2008hardware} implemented a PID controller combined with an evolutionary technique to tune control gains.
Generally,  classical controllers often lack optimality in achieving desired performances.
\cite{ellert1963synthesis} applied an optimization technique called the parametric expansion method to synthesize a linear time-varying aircraft landing controller. In \cite{liao2005fault,lungu2013automatic,niewoehner1994design,shue1999design,guan2018prescribed} issues such as robustness to measurement noise, actuator faults, wind disturbances, and flight uncertainties were investigated and $H_2, H_\infty$, LQR/LQG, mixed $H_2/H_\infty$ control techniques were employed to design optimal and robust aircraft landing controllers.
 Recently, concepts from adaptive control and intelligent control such as adaptive synthesis based on dynamic inversion theory \cite{lungu2013automatic,singh2008automatic,tripathi2016autonomous}, neural networks \cite{lau2007neural}, quantitative feedback theory \cite{wagner2007digital}, sliding mode control \cite{rao2014automatic}, structured singular value $\mu$-synthesis, or fuzzy techniques \cite{juang2005fuzzy} were used to design automatic aircraft landing controllers.  In \cite{prasad2007automatic,guan2018prescribed,rao2014automatic} nonlinear control techniques were applied to improve the control performance of the landing controller. 
Most of the aforementioned techniques are based on the asymptotic convergence of tracking errors. This does not guarantee a safe, smooth, and comfortable landing, which often requires the aircraft to accurately track a particular landing profile while landing within a given period of time and satisfying performance requirements and flight constraints.

The contribution of this paper is, therefore, designing an optimal controller that guarantees the finite-time convergence of the landing trajectory tracking errors. For this purpose, we first design the desired landing path which satisfies performance requirements and flight constraints. The flare-out maneuver phase is designed as an exponential path, which connects end point of glide slope phase and the touchdown point on the runway. Furthermore, the touchdown point is captured as a boundary condition. In addition, performance requirements and constraints on control input and output signals are determined. 
Subsequently, we formulate the tracking problem by designing a performance index which should be minimized in a targeted landing time. We then have converted the aircraft landing problem to a Linear Quadratic Tracking (LQT) problem where we minimize the designed cost function for a linear model of the aircraft respecting boundary conditions. Control gains are found by solving differential Riccati equations backward in time using the Dormand Prince algorithm. The simulation results are analyzed to show that the developed trajectory tracking system can handle a range of angle of attack and headwind disturbances.

  In the remainder of this paper, in Section \ref{sec:prob}, the aircraft landing problem is formulated.
In Section \ref{sec:proposed}, the desired trajectory and performance index are designed by incorporating the performance requirements and flight constraints. An LQT controller is developed for aircraft landing by minimizing the designed performance index to guide the aircraft to follow the desired landing trajectory. In Section \ref{sec:simu}, the proposed flight controller is simulated and the results are discussed. Finally, Section \ref{sec:concu} concludes this paper.

\section{Problem formulation}
\label{sec:prob}
\begin{figure}[h]
\includegraphics[scale=0.40]{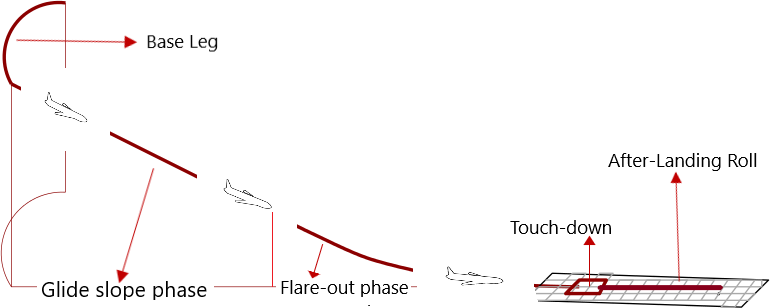}
\centering
\caption{An aircraft landing phases: the base leg, the glide slope, the flare-out, the touchdown, and the after-landing roll.}
\label{fig:lan}
\end{figure}

The aircraft landing process is described in Figure \ref{fig:lan}. The glide slope phase and flare-out phase are the most critical steps. After the base leg, the aircraft should descend towards the runway center-line with the flight path angle $\nu$ between 2.5 to 3.5 deg. As the aircraft descends to a certain altitude above the ground, the \textit {flare-out} phase begins. In the \textit {flare-out} phase of landing, the pitch angle $\theta$ of the airplane must be gradually adjusted to have a value between 0 to 10 deg, and the aircraft should approach the touchdown point  smoothly. The flare-out phase is the most difficult part of the landing maneuver, as the run-way is not in the pilot's field of view and the touchdown point has to be accurately reached within a targeted time smoothly, following the landing trajectory. In this paper, we assume that the lateral and longitudinal dynamics of the aircraft are decoupled, and the lateral controller \cite{lungu2016automatic} takes care of the lateral deviations from the center line of the runway.  
The focus of this paper, therefore, is on the design of a longitudinal controller to track the desired landing trajectory.
The longitudinal motion of the aircraft is governed entirely by the elevator deflection angle $\delta_e(t)$. The longitudinal dynamics of an aircraft is given by the so-called short period equation of motion \cite{ellert1963synthesis} of the aircraft:
\begin{equation}\label{eq2:model}
\dot{\theta}(s) = \frac{K_s(T_ss+1)}{(\frac{s^2}{\omega_s^2}+\frac{2\zeta s}{\omega_s}+1)}\delta_e(s)
\end{equation}
where $\dot{\theta}$ is pitch rate, $K_s$ is short period gain, $T_s$ is path time constant, $\omega_s$ is short period resonant frequency, and $\zeta$ is short period damping factor.
Assuming the velocity $V$ of the aircraft to be constant, pitch angle rate $\dot{\theta}$  and altitude $h$ are related \cite{ellert1963synthesis} through vertical acceleration $\ddot{h}$ as:
\begin{equation}\label{eq3:model2}
\ddot{h}(s) = \frac{V}{T_ss+1}\dot{\theta}(s)
\end{equation}
Combining (\ref{eq2:model}) and (\ref{eq3:model2}), a fourth-order linear differential equation is found:
\begin{equation}\label{eq3:model3}
h^{(4)}(t)+2\zeta\omega_s h^{(3)}(t)+\omega_s^2\ddot{h}(t) = K_sV\omega_s^2\delta_e(t)
\end{equation}
Substituting
\small
\begin{subequations}\label{eq4:relation}
\begin{align}
\ddot{h}(t) = \frac{V}{T_s}\theta(t) - \frac{1}{T_s}\dot{h}(t), h^{(3)}(t)) = \frac{V}{T_s}\dot{\theta}(t) - \frac{1}{T_s}\ddot{h}(t)\\h^{(4)}(t) = \frac{V}{T_s}\ddot{\theta}(t)- \frac{1}{T_s}h^{(3)}(t)
\end{align}
\end{subequations}
\normalsize
in (\ref{eq3:model3}), the new aircraft equation of motion becomes
\small
\begin{equation}\label{eq5:model4}
\begin{split}
\frac{V}{T_s}\ddot{\theta}(t) - (\frac{V}{T_s^2}-2\zeta\omega_s\frac{V}{T_s})\dot{\theta}(t) -
(2\zeta\omega_s\frac{V}{T_s^2}-\omega_s^2\frac{V}{T_s^2}\\-\frac{V}{T_s^3})\theta(t)-
(\frac{1}{T_s^3}-2\zeta\omega_s\frac{1}{T_s^2}+\frac{\omega_s^2}{T_s})\dot{h}(t) 
= K_sV\omega_s^2\delta_e(t)
\end{split}
\end{equation}
\normalsize
from which one can derive, the state space representation of the aircraft's model as:
\begin{subequations}
\label{eq:model}
\begin{align}
\dot{x}(t)= Ax(t)+B\delta_e(t) \\
y = Cx(t) 
\end{align}
\end{subequations}
where $ x(t) = \big[\begin{smallmatrix} h(t) & \dot{h}(t) & \theta(t) & \dot{\theta}(t)\end{smallmatrix}\big]^T$, $B =\big[\begin{smallmatrix} 0 & 0 & 0 & b_4 \end{smallmatrix}\big]^T$,\\ $A=\bigg[\begin{smallmatrix} 0 & 1 & 0 & 0\\0 & a_{22} & a_{23}&0\\ 0 & 0 & 0 & 1\\ 0 & a_{42} & a_{43} & a_{44}\end{smallmatrix}\bigg]$,
$a_{22} = \frac{-1}{T_s}$, $a_{23} = \frac{V}{T_s}$, $a_{42} = \frac{1}{vT_s^2}-2\frac{\zeta\omega_s}{vT_s}+\frac{\omega_s^2}{v}$, $a_{43} = 2\frac{\zeta\omega_s}{T_s}-\omega_s^2-\frac{1}{T_s^2}$, $a_{44} = \frac{1}{T_s}-2\zeta\omega_s$, $b_4 = \omega_s^2k_sT_s$ and C is the identity matrix.
\begin{figure}[t]
\includegraphics[scale=0.280]{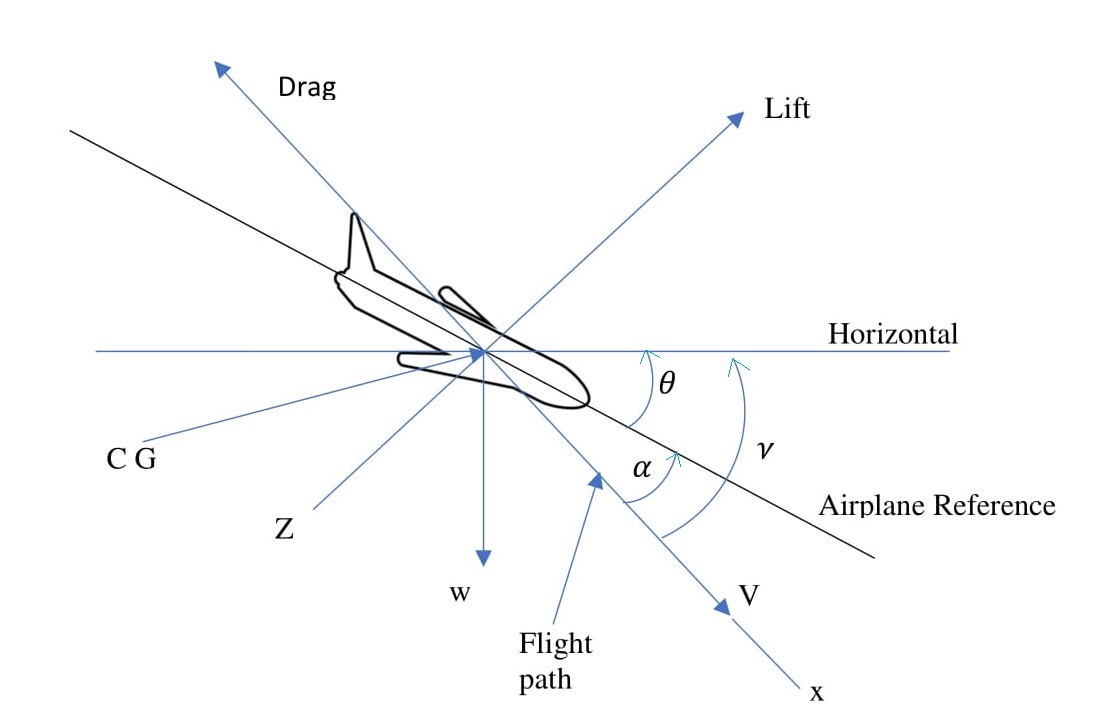}
\caption{Definition of aircraft coordinates and angles \cite{ellert1963synthesis}. $\alpha$ is the angle of attack, $\theta$ is the pitch angle, $\nu$ is the flight path angle.}
\label{fig:angles}
\end{figure}

Our aim is to design a controller for the aircraft to follow the landing trajectory with the minimum tracking error $e(t)$ given as:
\begin{equation}
e(t) = x(t)-r(t)
\end{equation}
where $r(t)$ $\in$ $\mathbb{R}^{4}$ is the desired landing trajectory over the landing time interval $[t_0$,  $t_f]$.
Then, given an aircraft model in (\ref{eq:model}) and  boundary conditions as $x(t_0) = x_0$ and $x(t_f) = x_f$, we address the following problems for the aircraft landing:
\begin{myprob}
Design a desired landing flight trajectory $r(t)$ which captures performance requirements and flight constraints.
\end{myprob}
\begin{myprob}
Design a performance index $J$ to capture  performance requirements and constraints
\end{myprob}
\begin{myprob}
Design a controller that tracks the desired trajectory $r(t)$ and satisfies the given performance requirements and constraints.
\end{myprob}
\section{Landing trajectory tracking}
\label{sec:proposed}
\subsection{Trajectory design}
\begin{figure}[h]
\includegraphics[scale=0.4]{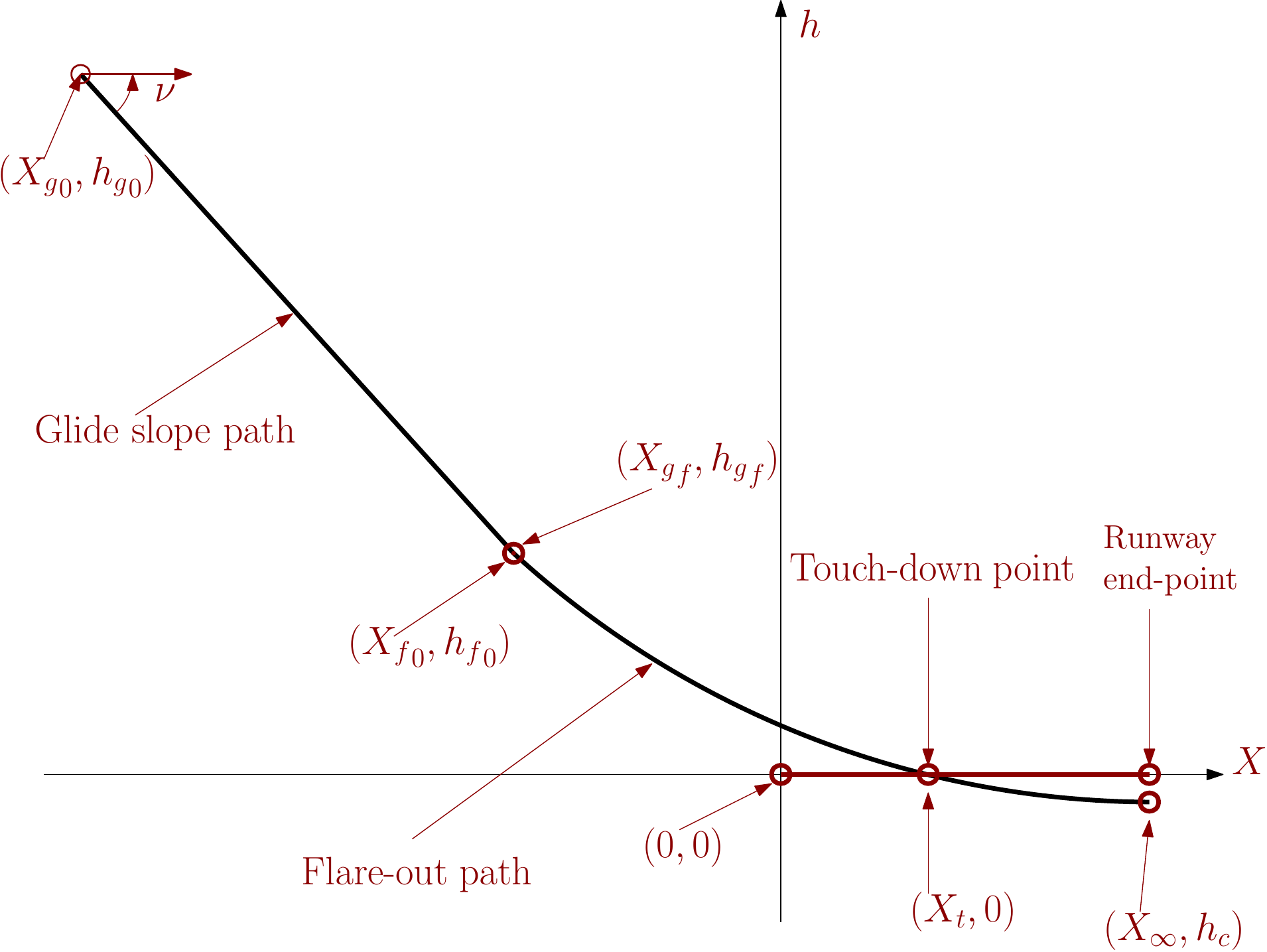}
\centering
\caption{An aircraft landing path where $({X_g}_0, {h_g}_0)$ and $({X_g}_f, {h_g}_f)$ are the beginning and the endpoint of the glide slope path; $({X_f}_0, {h_f}_0)$ and $(X_t, 0)$ are the beginning and the endpoint of the flare-out path; $(0,0)$ is the origin of X-h plane, and $(X_{\infty}, h_c)$ is the final point on the flare-out path usually defined below the ground. The origin of the  X-h plane is assumed to be at the beginning of the runway and the aircraft must land within one-third of the runway.}
\label{fig:aircraftlandingpath}
\end{figure}

The aircraft landing trajectory is the desired aircraft landing path, which the aircraft follows over a specified period of time $[t_0,t_f]$. Landing trajectory must ensure safe and comfortable landing. The desired landing path is expressed as a desired altitude which is a function of forward distance. Its important phases are the glide slope phase and flare-out phase as shown in Figure \ref{fig:aircraftlandingpath}. The glide slope phase is the first part of a landing path. It is modeled as a straight line path whose slope is given in terms of the flight path angle $\nu $.
 The other part is the flare-out path which is initiated when the aircraft reaches to a certain height ${h_f}_0$ above the landing surface. Flare-out path is an exponential path with a near zero flight path angle at touchdown. In Figure \ref{fig:aircraftlandingpath}, the desired landing path is shown in forward distance (X)-altitude (h)-plane. 

In the glide slope phase the aircraft tracks a straight line path pointing toward the beginning of the runway. The desired altitude, $h^d$, is computed based on a given flight path angle $\nu $:
\begin{equation}\label{eq:glide-slope}
h^d = -tan(\nu^d)(X-{X_g}_0)+{h_g}_0
\end{equation}
As soon as the flare-out phase starts, the desired altitude $h^d$ is computed as:
\begin{equation}
\label{eq:flare}
\begin{split}
h^d = -h_c+({h_{f}}_0+h_c)e^{-K_x(X - {X_{f}}_0)}
\end{split}
\end{equation}
where $K_x$ is a constant which defines the curvature of flare-out maneuver path, depending on the distance between the origin of X-h plane and the touchdown point.

The objective of a good, stabilized final approach is to descend at an angle and airspeed that permits the airplane to reach the desired touchdown point with minimum floating; in essence, a semi-stalled condition \cite{aviation2004airplane}. To accomplish this, it is essential that both the descent angle and the airspeed be accurately controlled. Since on a normal approach the power setting is not fixed as in a power-off approach, the power and pitch attitude are adjusted simultaneously as necessary to control the airspeed and the descent angle, or to attain the desired  altitudes along the approach path.
In this paper, we focus on the flare-out phase to achieve a safe and comfortable landing. 
In (\ref{eq:flare}), ${X_{f}}_0$, $h_c$ and $K_x$ are unknowns. We can solve for these unknowns from $tan(\nu^d)$, ${h_{f}}_0$, and $X_t$ under the following constraints:
\begin{itemize}
\item Slope continuity constraint: For a smooth phase change, the slope at the beginning of the flare-out path should be equal to the slope of the glide slope path.
Thus, the slope of glide slope path can be calculated as
\begin{equation}
\label{eq:slopeglide}
\dot{h^d} = -\tan(\nu^d)
\end{equation}
Similarly, the slope at the beginning of flare-out path is 
\begin{equation}
\label{eq:slopeconstr}
\dot{h^d} = {-K_x}({h_{f}}_0+h_c)e^{-K_x(X - {X_{f}}_0)}|_{X = {X_{f}}_0}
\end{equation}
Using (\ref{eq:slopeglide}) and (\ref{eq:slopeconstr}), $K_x$ is computed as
\begin{equation}
\label{eq:Kx}
K_x = \frac{\tan(\nu^d)}{{h_f}_0+h_c}
\end{equation}
\item Path continuity constraint: This constraint guarantees the continuity of the path when transiting from the glide slope phase to the flare-out phase. By design, we have ${X_f}_0={X_g}_f$. We then need to ensure that $h^d$ at ${X_g}_f$ given in (\ref{eq:glide-slope}) is the same as  $h^d$ at  ${X_f}_0$ given in (\ref{eq:flare}) resulting in
\begin{equation}
\label{eq:pathcontinuity}
{X_{f}}_0 = \frac{{h_{g}}_0-{h_{f}}_0}{tan(\nu^d)}+{X_{g}}_0 \\
\end{equation}
\item Touchdown constraint: The landing path intersects the ground at the touchdown point $X_t$. Thus, at touchdown point we have
\begin{equation}
\label{eq:touchdown}
 -h_c+({h_{f}}_0+h_c)e^{-K_x(X_t - {X_{f}}_0)}=0
\end{equation}
Finally, $K_x$ and $h_c$ are found by solving (\ref{eq:Kx}), (\ref{eq:pathcontinuity}) and (\ref{eq:touchdown}).
\end{itemize}
Under the assumption of constant aircraft forward velocity $\dot{X}$, at any time moment $t > t_0$, the forward distance $X$ on the flare-out curve will be approximately
\begin{equation}
\label{eq:constvel}
X = {X_{f}}_0+\dot{X}(t-t_0)
\end{equation}
Substituting $X-{X_f}_0$  from  (\ref{eq:constvel}) in (\ref{eq:flare}), the landing trajectory in flare-out phase is governed by
\begin{equation}
\label{eq:altitude}
h^d = -h_c+({h_{f}}_0+h_c)e^{-K(t-t_0)}
\end{equation}
where $K = K_x\dot{X}$.
By differentiating (\ref{eq:altitude}), the desired rate of descent in flare-out phase is:
\begin{equation}
\dot{h_d} = -K({h_{f}}_0+h_c)e^{-K(t-t_0)}
\end{equation}
The value of the pitch angle is crucial only during the last few sec of landing before the touchdown point.
In flare-out maneuver phase, the desired pitch angle,  $\theta_d$, is::
\begin{equation}
\theta_d = 0
\end{equation} 
Also, for a safe and comfortable landing, the pitch rate $\dot{\theta_d}$ should be:
\begin{equation}
\dot{\theta_d} = 0
\end{equation}

\subsection{Performance requirements}
\label{sec:perfo}
The following landing requirements and flight constraints are considered for the aircraft landing in this paper:
\begin{enumerate}
\item [C1:] The landing path should be an exponential path such that it ensures a safe and comfortable landing.
\item [C2:]The rate of descent $\dot{h}$ should be non-zero in order to avoid overshooting. In addition, the rate of descent at touchdown should be a small negative value in order to prevent over stressing of landing gear. Requirements to avoid undesirable behaviors such as landing gear over stress, aircraft floating down runway and aircraft overshooting runway are captured by magnitude of rate of descent at touchdown point. Typically based on type of the aircraft, a value between 60 to 180 ft/min is considered as an ideal rate of descent at touchdown point. 
\item [C3:] The lower limit  on  the pitch angle $\theta$ should be $0^{\circ}$ in order to prevent the nose wheel of an aircraft from touching down first, and the upper limit on the pitch angle $\theta$ should be $10^{\circ}$ to prevent the tail gear from touching down first:
\item [C4:]During landing, in the flare-out maneuver phase, the angle of attack must remain below 80 percent of the stall value. The stall value is assumed to be $18^{\circ}$ \cite{ellert1963synthesis}. Hence, the rate of change of angle attack is restricted to a value less than 20 percent of the stall value:
\item [C5:]The longitudinal motion control is mainly executed by the elevator. To avoid saturation, the elevator is not permitted to operate against the mechanical stops during the landing process. Thus, the deflection of the elevator is restricted between $-35^{\circ}$ and $+15^{\circ}$.
\end{enumerate}
\normalsize
In addition, the aircraft should follow the designed landing trajectory and reduce the tracking error within the targeted landing time $t_f$ with minimum control effort $\delta_e(t)$. This can be captured by the following  performance index:
\normalsize

\vspace{-0.3cm}
\small
\begin{equation}
\label{eq:cost}
\begin{split}
J = [Cx(t_f)-r(t_f)]^TP[Cx(t_f)-r(t_f)]+\\
\int_{t_0}^{t_f}\{[Cx(\tau)-r(\tau)]^TQ[Cx(\tau)-r(\tau)]+\delta_e^T(\tau)R\delta_e(\tau)\}d\tau
\end{split}
\end{equation}
\normalsize
where $P \geq 0$, $Q \geq 0$, $R>0$,  are all symmetric weighting parameters.
The terms outside the integral ensure that the states are close to the desired values at the touchdown point, avoiding an early or  a late landing. On the other hand, the terms inside the integral push the aircraft to track the desired landing trajectory with minimum control effort  in the landing time $[t_0,t_f]$.

\subsection{Controller design based on linear quadratic tracking}
Given  the aircraft model in (\ref{eq:model}) and the boundary conditions $ x(t_0) = x_0 =\big[\begin{smallmatrix} h(t_0)={h_f}_0 & \dot{h}(t_0) & \theta(t_0) & \dot{\theta}(t_0)\end{smallmatrix}\big]^T$ and  $x(t_f)=x_f =\big[\begin{smallmatrix} 0 & 0 & 0 & 0 \end{smallmatrix}\big]^T$, our aim is to find an optimal control $\delta^*_e(t)$ that guides the aircraft to follow the desired landing trajectory $r(t)$ over the time interval  $[t_0,t_f]$, minimizing  the performance index in (\ref{eq:cost}), as described below:
\begin{equation}\label{constrainedoptimization}
\min\limits_{\forall t\in [t_0 \,\, t_f]} J
\end{equation}
\noindent subject to
\begin{equation*}
\begin{cases}
\dot{x}(t) = Ax(t)+B\delta_e(t)\\
x(t_0) = x_0\\
x(t_f) = x_f
\end{cases}
\end{equation*}
Defining the co-state vector \small $\lambda(t)=S(t)x(t)-v(t)$\normalsize, where \small $S(t)= C^TP(t)C(t)$ \normalsize and \small$v(t)=C^TPr(t)$\normalsize, and defining the Hamiltonian function \small$H=\frac{1}{2}[Cx(t)-r(t)]^TQ[Cx(t)-r(t)]+\frac{1}{2}\delta_e^T(t)R\delta_e(t)+\lambda^T(Ax(t)+B\delta_e(t))$\normalsize, the constrained optimization in (\ref{constrainedoptimization}) will be converted to
\small
\begin{equation}\label{unconstrainedoptimization}
\begin{split}
\min\limits_{\forall t\in [t_0 \,\, t_f]} \bar{J}=[Cx(t_f)-r(t_f)]^TP[Cx(t_f)-r(t_f)]+\\
\int_{t_0}^{t_f}(H-\lambda^T\dot{x}(t))d\tau
\end{split}
\end{equation}
\normalsize
whose solution can be found from $\delta J(x(t))=0$, where  $\delta J$ is the variation of the functional $J$. The optimal control function will be
\small
\begin{subequations}
\begin{align}
\delta^*_e(t)=-K(t)x(t)+R^{-1}B^Tv(t)\\
K(t) = R^{-1}B^TS(t)
\end{align}
\end{subequations}
\normalsize
in which $S(t)$ and $v(t)$ can be found by solving the following differential equations in a backward way with a given boundary conditions:
\small
\begin{subequations}
\begin{align}
-\dot{S} = A^TS+SA-SBR^{-1}B^TS+C^TQC \\
-\dot{v} = (A-BK)^Tv+C^TQr \\
S(t_f) = C^TPC \\
v(t_f) = C^TPr(t_f)
\label{eq:DiffEq}
\end{align}
\end{subequations}
\normalsize

\section{Simulation}
\label{sec:simu}
A simulation environment is set up by integrating the MATLAB and FlightGear simulation Software in which we simulated an aircraft landing from the altitude of 95 ft to the runway 10L of San Francisco airport. The information on the approach plate (see Table \ref{table:approachplateparameters}) of runway 10L of the airport is used to design a safe and comfortable landing trajectory for an aircraft with model parameters given in Table \ref{table:aircraftparameters}. 
 \begin{table}[b]
 \parbox{.45\linewidth}{
 
 \centering
 \begin{tabular}[t]{ || p{0.25cm} | p{0.75cm} | p{1.5cm}|}
 \hline
$ N_0$  & Symbol & Parameter value \\
 \hline\hline
 1 &  $K_s$  & -0.95 $sec^{-1}$ \\
 2 &  $T_s$ & 40 $sec$  \\
 3 &  $\omega_s$ & 1 $rad/sec$ \\
 4 &  $\zeta$ & 0.5 \\
 \hline
 \end{tabular}
 \caption{The aircraft model parameters}
\label{table:aircraftparameters}
}
\hfill
\parbox{.55\linewidth}{
\small
\vspace{-0.98cm}
\begin{equation}
\begin{aligned}
\begin{split}
P = \bigg[\begin{smallmatrix} 0.9 & 0 &0 &0\\
0 & 0.01&0&0\\
0&0&1&0\\
0&0&0&1
\end{smallmatrix}\bigg] \\
Q = \bigg[\begin{smallmatrix} 0.00067 & 0 &0 &0\\
0 & 0.0265&0&0\\
0&0&150&0\\
0&0&0&65
\end{smallmatrix}\bigg]\\
R = 1
\end{split}
\end{aligned}
\label{eq:param}
\end{equation}
\normalsize
}
\end{table}
Table \ref{table:approachplateparameters} summarizes parameters extracted from the approach plate. Similarly, assuming this particular aircraft has a nominal initial altitude ${h_f}_0$ of 100 ft, the desired flight path angle $\nu$ of 3 deg, and the pitch angle $\theta$ of 0 deg, then using (\ref{eq:Kx}) to (\ref{eq:constvel}), we found the desired landing trajectory parameters ${X_f}_0$, $K_x$, $h_c$, $K$, as shown in Table \ref{table:Landingparameters}. 

\begin{figure}[h]
\includegraphics[scale=0.6]{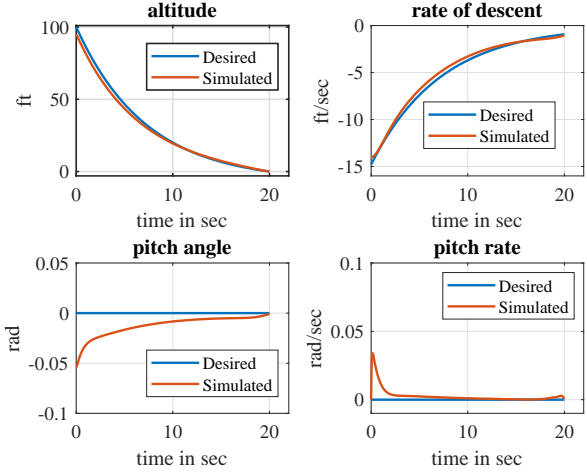}
\caption{The desired smooth landing trajectories (solid blue) and  the simulated aircraft landing  trajectories (solid red).}
\label{fig:trackingresult}
\end{figure}

The designed desired landing trajectory is plotted in Figure \ref{fig:trackingresult} in solid blue. The desired altitude $h^d(t)$ and the desired altitude rate $\dot{h}^d(t)$  are exponential functions while the desired pitch angle $\theta_d$ and the desired pitch rate $\dot{\theta}_d$  are constant functions. Both exponential functions and constant functions have derivatives of all orders. Therefore, the designed landing trajectory is a smooth landing trajectory.
 \begin{table}[t]
 \parbox[t]{.47\linewidth}{
\vspace{0.2cm}
 \begin{tabular}{ || p{0.25cm} | p{0.5cm} | p{1.33cm}|}
 \hline
$ N_0$ & Name & Parameter value \\
 \hline\hline
 1 & ${X_g}_0$  & -34346 $ft$ \\
 2 & ${h_g}_0$ & 1800 $ft$  \\
 3 & $X_t$ & 3957 $ft$ \\
  \hline
 \end{tabular}
 \caption{ Parameters extracted from the approach plate}
\label{table:approachplateparameters}
\vspace{0.75cm}
 \begin{tabular}{ || p{0.25cm} | p{0.5cm} | p{1.33cm}|}
 \hline
$N_0$ & Name & Parameter value \\
 \hline\hline
 1 & $K_x$  & 0.00049 \\
 2 & $\dot{X}$ & 256 $ft/sec$  \\
 3 & $K$ & 0.1385  \\
 4 & $h_c$ & 6.68 $ft$ \\
 5 & ${X_f}_0$ & -1908 $ft$ \\
 \hline
 \end{tabular}
 \caption{The designed landing trajectory parameters}
\label{table:Landingparameters}
}
\hfill
\parbox[t]{.45\linewidth}{
\vspace{0.2cm}
 \begin{tabular}[t]{||p{0.25cm} | p{1.1cm}| p{1.1cm}|}
 \hline
 $N_0$ & Parameter Name & Parameter value \\ [0.5ex]
 \hline\hline
 1 & $t_0$ & 0 sec \\
 2 & $t_f$ & 20 sec  \\
 3 & $h_0$ & 95 ft \\
 4 & $\dot{h}_0$ & -14 ft/sec \\
 5 & $\theta_0$ & -0.05 rad \\
 6 & $\dot{\theta}_0$ & 0 rad/sec \\
 7 & $h_f$ & 0 ft \\
 8 & $\dot{h}_f$ & 0 ft/sec \\
 9 & $\theta_f$ & 0 rad \\
 10 & $\dot{\theta}_f$ & 0 rad/sec \\[1ex]
 \hline
 \end{tabular}
 \caption{Simulation parameters: Initial and final conditions for case I}
\label{table:simulationparameters}
}
\end{table}
%

\subsection{Case I: Simulation results for the  initial condition given in Table \ref{table:simulationparameters}} 
For the given initial conditions in Table \ref{table:simulationparameters}, the simulated aircraft landing trajectory is shown in Figure \ref{fig:trackingresult} in red. The controller gain is found solving (\ref{eq:DiffEq}) in a backward way using the Dormand-Prince algorithm. To respect the Constrains C1-C5 in Section \ref{sec:perfo},  we have tuned the performance index parameters $R$, $Q$, and $P$ to values in (\ref{eq:param}).
Comparing the desired altitude $h^d(t)$, the desired altitude rate $\dot{h}^d(t)$, and their respective simulation values in Figure \ref{fig:trackingresult}, we can see that the aircraft has accurately tracked the desired landing trajectory in a desired landing time of 20 sec satisfying Constraints C1-C5 in Section \ref{sec:perfo}. This comparison is summarized in Table \ref{table:constraintvalidation}.
\begin{figure}[h]
\centering
\includegraphics[scale=0.5]{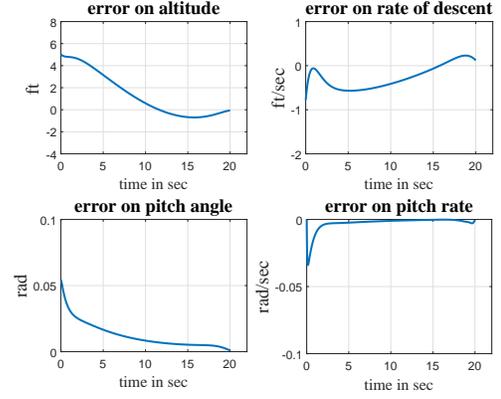}
\caption{The trajectory tracking error for the flight simulation results in Fig. \ref{fig:trackingresult}}
\label{fig:error}
\end{figure}
\begin{figure}[h]
\centering
\includegraphics[scale=0.5]{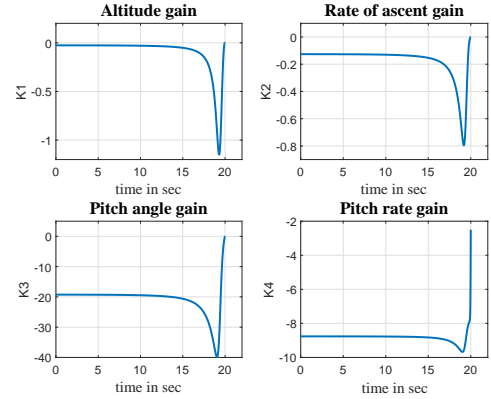}
\caption{Optimal controller gains for the flight simulation results in Fig. \ref{fig:trackingresult}}
\label{fig:controlgain}
\end{figure}
 \begin{table}[t]
\vspace{0.25cm}
 \begin{tabular}{ || p{0.25cm} |  p{3.67cm} | p{3.1cm}|}
 \hline
$N_0$ & Constraint &Constraint validation based on simulated flight data \\
 \hline\hline
 C1 & Exponential (smooth) trajectory  & Satisfied by design in (\ref{eq:flare})\\
 \hline
 C2 & $60\leq\lVert\dot{h}^d(t_f)\rVert\leq 180ft/min$  & $\lVert\dot{h}^d(t_f)\rVert$ = 62.7 ft/min  \\
\hline
 C3 & $0^{\circ}\leq\theta(t_f)\leq+10^{\circ}$ & $\theta(t_f) \approx 0$  \\
 \hline
 C4 & \shortstack{$\alpha(t) < 18^{\circ}$\\ $\Delta\alpha(t) < 3.6^{\circ}$} & \shortstack{Satisfied through\\ $\theta \approx 0$ and $\dot{\theta} \approx 0$}  \\
     \hline
 C5 & $-35^{\circ}\leq\delta_e(t) \leq +15^{\circ}$ & $-22.3^{\circ}\leq\delta_e(t) \leq +2.4^{\circ}$ \\
 \hline
 \end{tabular}
  \caption {Landing constraints C1-C5 validation for the given initial conditions in Table \ref{table:simulationparameters}}
\label{table:constraintvalidation}
\end{table}
\subsection{Case II: Simulation results for different initial conditions}
\begin{figure}[h!]

\centering
\includegraphics[scale= 0.450]{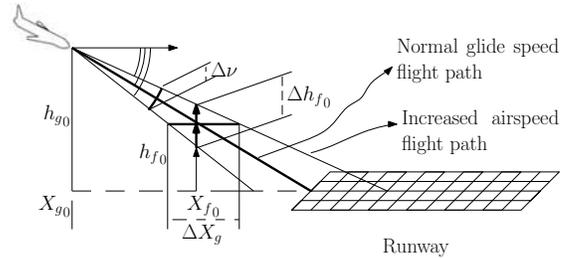}
\caption{The deviations in the gliding distance and the pitch angle due to head wind disturbances cause deviations in the initial altitude of the flare-out phase.}
\label{fig:generalcase}
\vspace{-0.7cm}
\end{figure}

Head wind and deviation of a desired flight path angle from a desired value of 3 deg introduces a deviation in the gliding distance and the pitch angle which in turn give rise to a deviation in the initial altitude of the flare-out phase as shown in Figure \ref{fig:generalcase}.
However, the designed controller can handle these deviations in the initial altitude $\Delta {h_f}_0$ and pitch angle  $\Delta \theta$ by adjusting control input in order to maintain the rate of descent and the desired approach airspeed, while avoiding the saturation region of the elevator actuator (Constraint C5) . 
\begin{figure}[h]
\centering
\includegraphics[scale= 0.6]{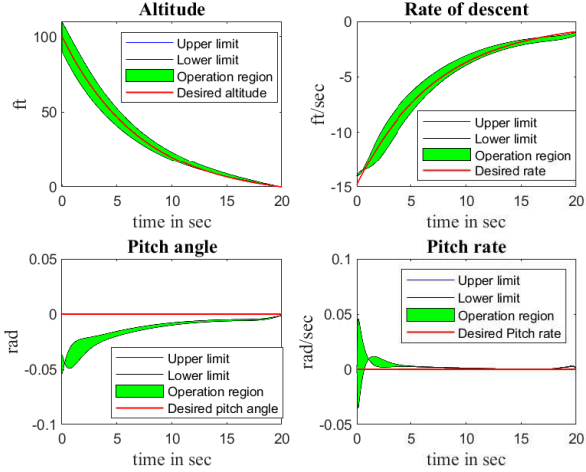}
\caption{Trajectory tracking performance for different initial conditions of the flare-out phase.}
\label{fig:linearoperation}
\end{figure}
Acceptable deviations in the initial altitude and pitch angle around the desired landing trajectory is determined by varying them correlationally until the control limiting values are attained instantaneously. Accordingly, $\Delta {h_f}_0$ and $\Delta \theta$ are found to be 20 ft and 1 deg respectively, and  the corresponding region of operation is shown in Figure \ref{fig:linearoperation} in green.
\begin{figure}[h!]
\begin{minipage}{.25\textwidth}
\includegraphics[scale=0.3]{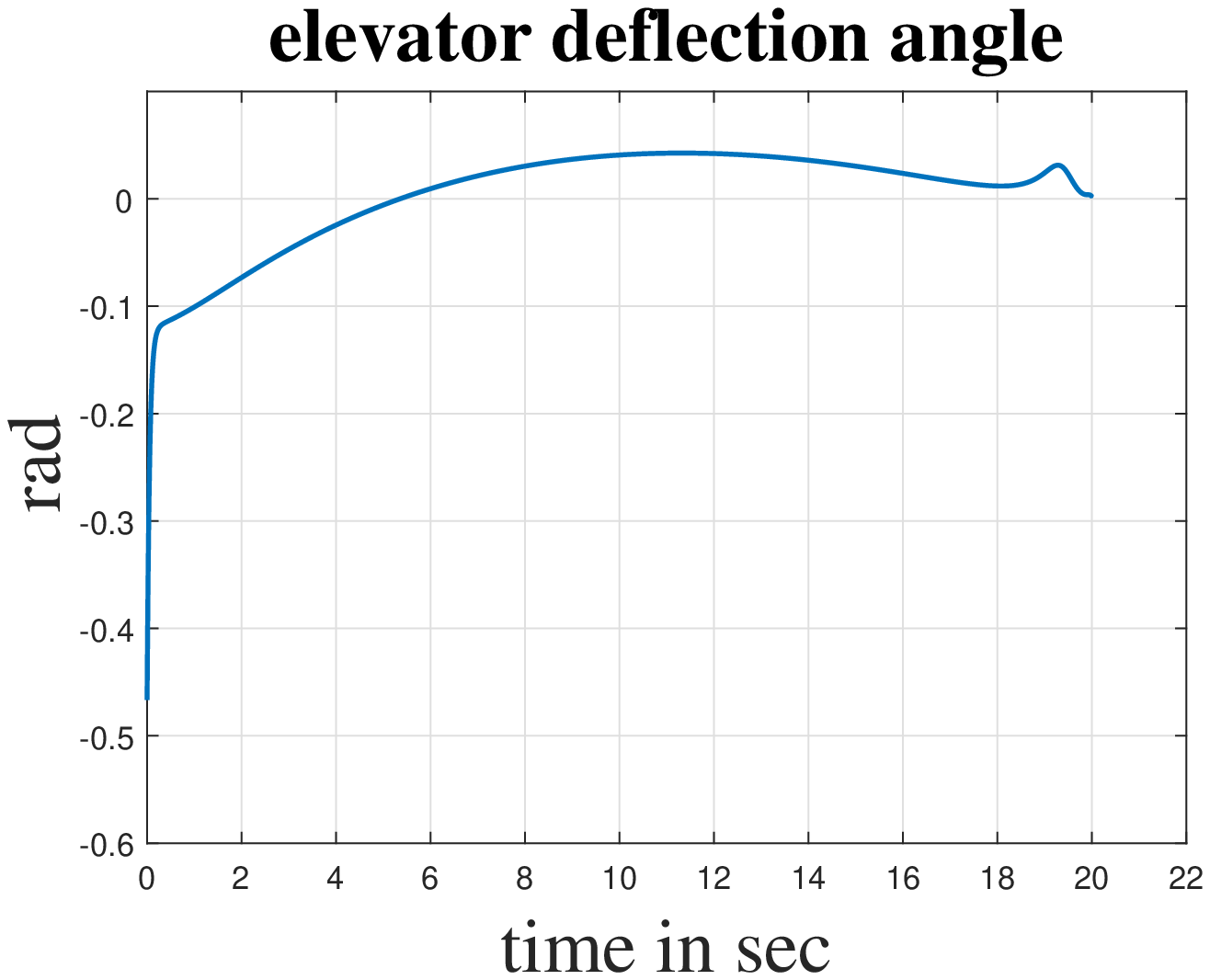}
\label{fig:controllimit}
\end{minipage}%
\begin{minipage}{.25\textwidth}
\includegraphics[scale= 0.3]{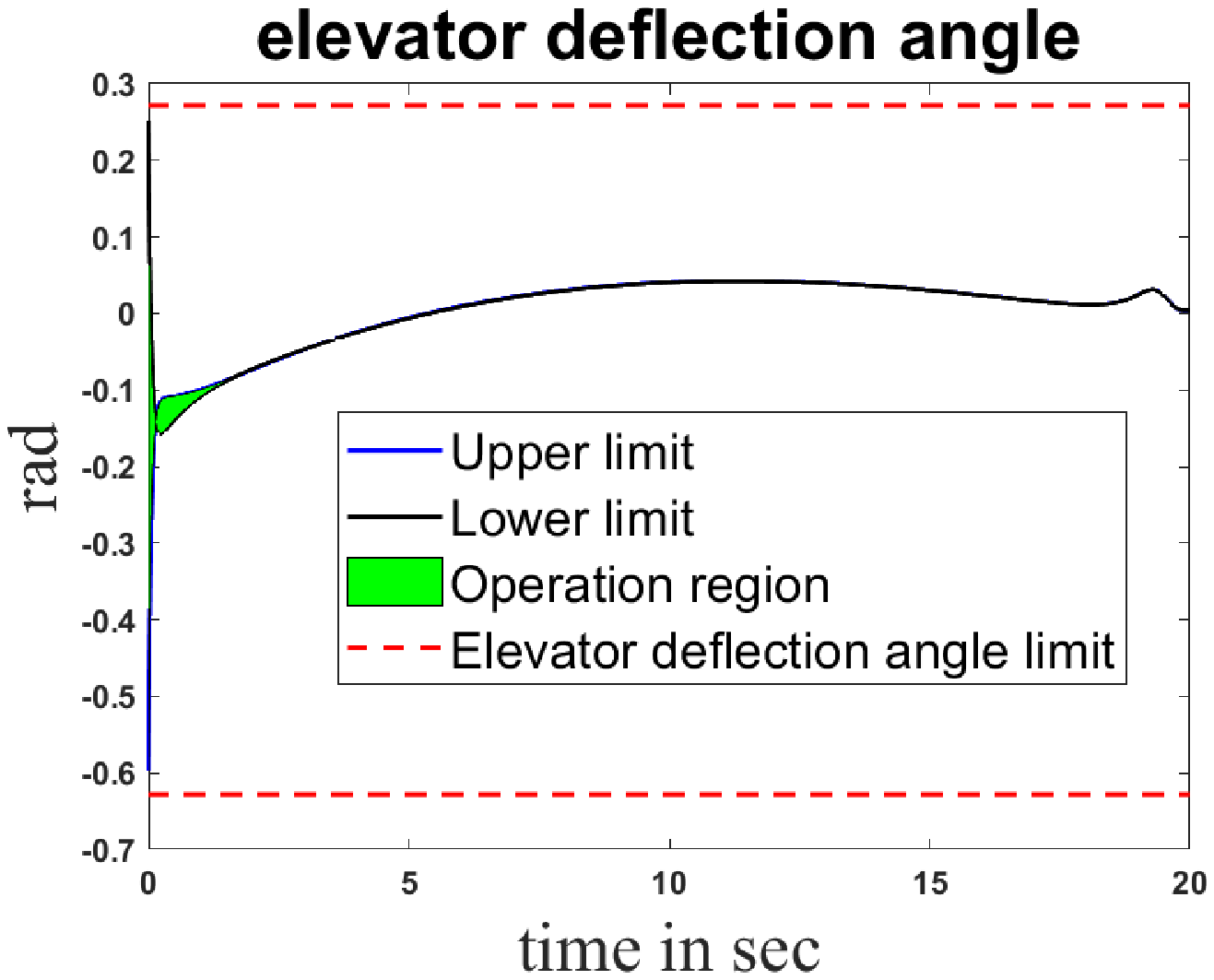}
\label{fig:elevatorangle}
\end{minipage}
\caption{The control input signal for the flight simulation results for case I (left) and case II (right).}
\end{figure}
\section{conclusion}
\label{sec:concu}
In this paper, a smooth trajectory tracking system was designed for a fixed-wing aircraft landing. The aircraft landing problem was converted to a finite-time LQT problem. Information on the approach plate of the runway was employed to design a desired smooth landing trajectory. We considered multiple performance requirements and constraints to design an optimal controller which minimizes the trajectory tracking error. The designed controller gives an accurate tracking performance for the flare-out phase of aircraft landing. The simulation result demonstrated the finite-time convergence of the trajectory tracking error. Moreover, the robustness of the designed controller was evaluated against different initial conditions due to the headwind disturbances. This work can be extended by employing a nonlinear model of an aircraft in the presence of turbulent crosswinds and system uncertainties.   
\section*{Acknowledgment}
This research is supported by Air Force Research Laboratory and OSD under agreement number FA8750-15-2-0116.
\bibliographystyle{IEEEtran}
\bibliography{ifacconf}
\end{document}